\documentclass[a4paper,11pt]{article}
\usepackage[toc,page]{appendix}

\usepackage[english]{babel}
\usepackage[utf8x]{inputenc}
\usepackage[T1]{fontenc}
\usepackage{authblk}

\usepackage[a4paper,top=3cm,bottom=2cm,left=3cm,right=3cm,marginparwidth=1.75cm]{geometry}

\usepackage{amsmath}
\usepackage{graphicx}
\usepackage{subcaption}
\usepackage{color} 
\usepackage{threeparttable}
\usepackage{dcolumn}
\usepackage{lmodern}

\usepackage[colorinlistoftodos]{todonotes}
\usepackage[colorlinks=true, allcolors=blue]{hyperref}

\title{Online Red Packets: A Large-scale Empirical Study
of Gift Giving on WeChat 
}

\author{Yuan Yuan\thanks{Institute for Data, Systems, and Society, Massachusetts Institute of Technology, Cambridge, MA, USA, 02139. yuan2@mit.edu.}, Tracy Xiao Liu\thanks{Department of Economics, School of Economics and Management, Tsinghua University, Beijing, China, 100084. liuxiao@sem.tsinghua.edu.cn.}, Chenhao Tan\thanks{Department of Computer Science, University of Colorado, Boulder, CO, USA, 80309. chenhao.tan@colorado.edu.}, and Jie Tang\thanks{Department of Computer Science and Technology, Tsinghua University, Beijing, China, 100084. jietang@tsinghua.edu.cn.}}

\begin{document}
\maketitle

\begin{abstract}
Gift giving is a ubiquitous social phenomenon, and
red packets have been used as monetary gifts in Asian countries for thousands of years.
In recent years, online red packets have become widespread in China through the WeChat platform.
Exploiting a unique dataset consisting of 61 million group red packets and seven million users, 
we conduct a large-scale, data-driven study to understand the spread of red packets and the effect of red packets on group activity.
We find that the cash flows between provinces are largely consistent with provincial GDP rankings, e.g., red packets are sent from users in the south to those in the north. 
By distinguishing spontaneous from reciprocal red packets, we reveal the behavioral patterns in sending red packets: males, seniors, and people with more in-group friends are more inclined to spontaneously send red packets, while red packets from females, youths, and people with less in-group friends are more reciprocal. 
Furthermore, we use propensity score matching to study the external effects of red packets on group dynamics.
We show that red packets increase group participation and strengthen in-group relationships, which partly explain the benefits and motivations for sending red packets.
\end{abstract}

\section{Introduction}

Online red packet (formerly called ``Lucky Money'' or ``Red Envelopes'')  was first released in 2014 as a new feature on WeChat, the largest social messaging platform in China. 
This feature enables people to send red packets to individual friends individually or randomly split a red packet among a limited number of people in a group on a first-come, first-serve basis.
WeChat red packets have become viral since 2015:
there were over 100 million participants during the 2016 Mid-autumn Festival alone, and the number of red packets sent over WeChat on Lunar New Year's Eve 2017 was 14.2 billion.
Now, to facilitate paricipation by users overseas, it is possible for someone to send red packets by linking her PayPal account to WeChat.
\footnote{The Economist offers an introduction to  WeChat and red packets
(\href{https://www.economist.com/news/business/21703428-chinas-wechat-shows-way-social-medias-future-wechats-world}{https://www.economist.com/news/business/21703428-chinas-wechat-shows-way-social-medias-future-wechats-world}). }

The red packet phenomenon 
is an instance of gift giving,
which is a pervasive behavior in human society that ranges from giving Christmas gifts to purchasing from a wedding registry \cite{cheal2015gift,mauss2000gift,sherry1983gift,caplow1984rule,akerlof1984gift,schwartz1967social,bendapudi1996enhancing}.  
However, gift giving is  governed almost entirely by unwritten rules.
Such unwritten rules have attracted interest from a wide range of disciplines, including anthropology, economics, sociology,  psychology, and marketing \cite{cheal2015gift,mauss2000gift,sherry1983gift,caplow1984rule,bendapudi1996enhancing}. 
Findings indicate that the motivations for gift giving include altruism \cite{andreoni2003charitable,midlarsky1989generous}, reciprocity \cite{komter1997gift,komter1996reciprocity}, and many other factors such as  prestige \cite{harbaugh1998donations}, empathy \cite{berger1962conditioning}, etc. 
For example, the red packets between friends are mainly reciprocal and it is considered as ``Renqing (favor)'' in Chinese culture \cite{chan2003art}. 
By contrast, sending red packets among family members is expected to enhance family bonds is usually considered non-reciprocal.

Our contribution to the literature is twofold. 
First, we study a new type of online red packets, the ``group red packet''.
In contrast with traditional red packets exchanged among family members and friends, this feature allows individuals to send red packets to 
a group of people by randomly splitting a red packet among a limited, user-determined number of people 
on a first-come, first-served basis. 
Moreover, online platforms provide a convenient and informal channel for gift giving, which may bring fundamental changes to the age-old practice of gift giving.
Second, most studies in the literature use small samples, including
studies of online gifts \cite{taylor2002age,suhonen2010everyday}. 
To the best of our knowledge, this is the first study to quantitatively analyze gift exchange using a dataset on millions of users.
In addition, considering a different setting than the fruitful existing empirical works focusing on the West \cite{andreoni2003charitable,kottasz2004differences,mesch2011gender,midlarsky1989generous,hansler1989geo,radley1995charitable,mears1992understanding,amato1987family,jones1991charitable,berg1995trust,komter1997gift,komter1996reciprocity,wang1997does,gneezy2006putting,plickert2007s,falk2007gift}, our study includes a large sample of the population in contemporary China and sheds light on gift giving in a representative Eastern country.

In this paper, we study this new type of gift giving behavior in three respects.
First, we provide an overview of the cash flow patterns
among users with different demographic and geographic backgrounds. 
If gift exchanges were completely reciprocal, the cash flows among users from different groups would be approximately zero. 
However, our results reveal intriguing patterns among different demographic and geographic groups. For example, 
we find that users from southern provinces are inclined to export cash to their northern counterparts.

Second, we study how demographic factors such as gender and age relate to sending red packets.
In particular, to distinguish the motivations for sending red packets 
(altruism vs. reciprocity), we define spontaneous and reciprocal red packets and 
identify their corresponding distinctive patterns.
We find that males, older people, southerners, and people with more friends in the group tend to send more spontaneous red packets, e.g., they are more likely to be the first one to send a red packet in a group, while red packets from females, younger people, northerners, and people with more friends in the group are more reciprocal. These differences can be explained by many aspects of Chinese culture, including ``mianzi'', i.e. the Chinese cultural value of interpersonal  dignity or prestige. 

Finally, we  examine the casual effect of online red packets on group dynamics.
Applying propensity score matching to data on approximately seven million red packets, we conduct a quasi-experiment to study the causal effects of the cash amount in red packets. 
We show that increasing the amount of money in red packets can stimulate reciprocal red packets and, consequently, increase the total amount of red packets sent by members of a group. 
We also find that increasing the amount of money in red packets can provide the senders with new friends. 
Based on these findings, we shed light on the benefits of online group red packets for both senders and all group members and improve our understanding of the motivation for sending red packets. 

\section{Results}
\subsection{Macro Cash Flow Patterns}

\begin{figure}
\centering
\includegraphics[width=\linewidth]{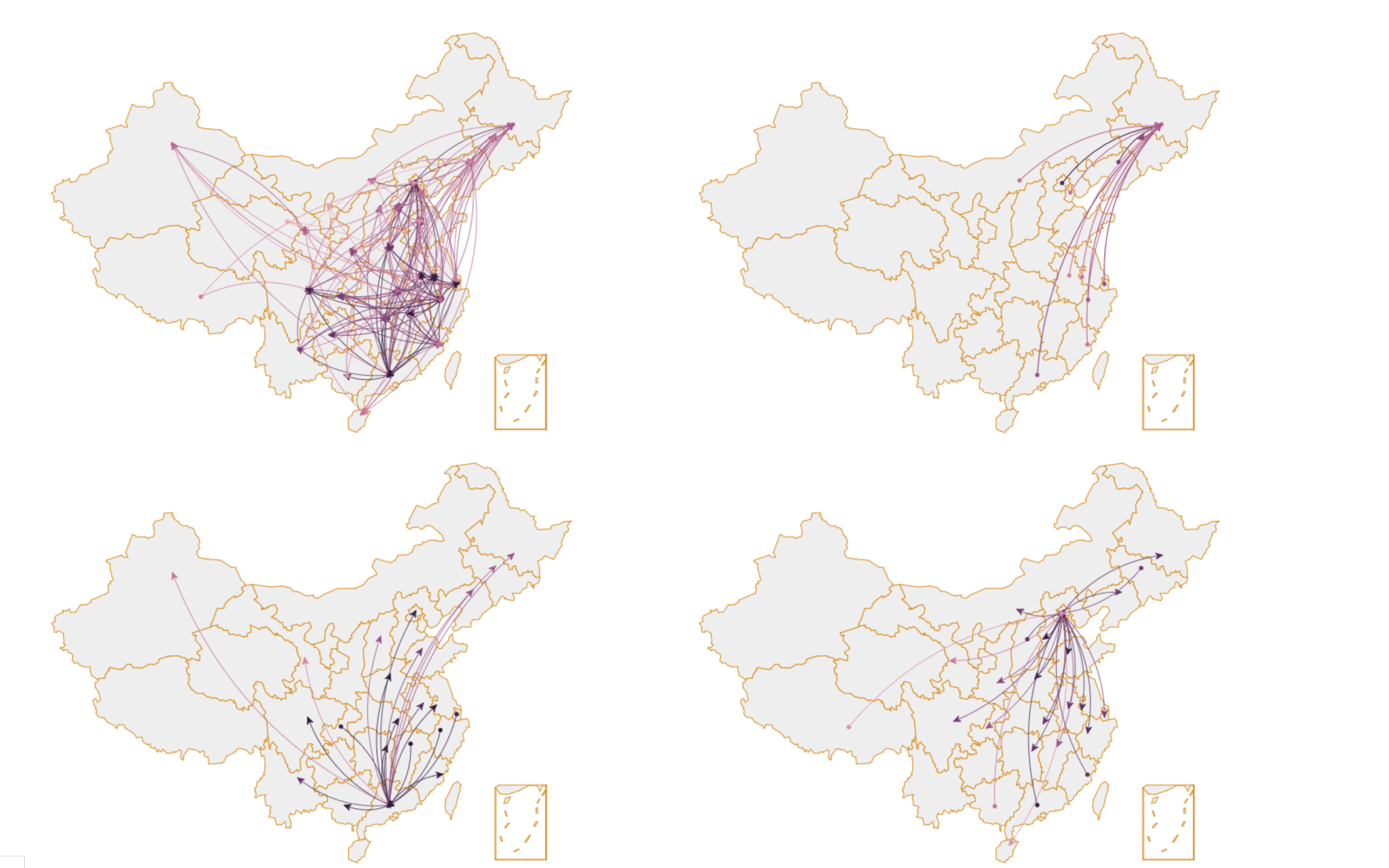}
\caption{The net cash flows between provinces in Mainland China (a: upper left); the cash flow for Heilongjiang Province (b: upper right); Guangdong Province (c: lower left); and Beijing Municipality (d: lower right). Each curve represents a net cash flow from the province at the origin of the line to the province at the arrow, and its darkness represents the size of net cash flow. 
}
\label{fig:provflow}
\end{figure}

We measure the directions of cash flows between users 
with different demographics. 
In particular, we are interested in examining whether, on net, the group red packets are sent from those from more-developed provinces to those from less-developed provinces. 

Figure~\ref{fig:provflow} (a) presents the net cash flow in Mainland China.\footnote{For easy visualization, we filter pairs of provinces where the total cash flow is under 1,000 RMB in the dataset.} First, we observe more net cash flow from the east coast to the west, or from the south to the north. This suggests that the direction of the cash flow might be significantly affected by the GDP in a region. Second, the magnitude of cash flow varies and the largest cash flow is from  Guangdong to Hunan. We conjecture that this is driven by the large number of migrants from Hunan to Guangdong. 

Furthermore, we are interested in understanding the relationship between the magnitude of cash flows and the geographic distance between two regions. We selected (1) Beijing, which is the capital of China and a municipality in the north with a large number of migrants, (2) Heilongjiang, which is a province in the northeast that has been experiencing significant economic difficulties and labor force outflows, and (3) Guangdong, which is the province with the highest GDP, population, and number of migrants. 

We find that the net cash flows from Heilongjiang to other provinces are almost always negative and the figures from Guangdong to other regions are almost always positive. Furthermore, the cash flows for both provinces are mainly with neighboring regions, and geographic distance tends to have a negative impact on the level of cash flows. 
In contrast, we do not find a consistent direction of cash flows for Beijing. 
On the one hand, the average standard of living in Beijing is higher than in most other provinces, leading to substantial cash flows out of this region; on the other hand, there are millions of college students in Beijing, and this may lead to large cash flows into the region. 
The cash flows between two arbitrary provinces and the net amount of incoming cash for each province are presented in SI.

To validate the aforementioned factors affecting the interprovincial cash flows, we predict the net cash flow between two provinces by the differences in GDP and GDP per capita (PPP) and the distance between their capitals. 
We find that a one-thousand-RMB increase in the difference in PPP is associated with a $9.02$ increase in the net cash flow ($p<0.01$), while a one-kilometer increase in distance is associated with a $0.57$ decrease in the net cash flow ($p<0.01$), while the results for GDP are not significant. 
The difference in PPP measures the difference in standards of living between provinces, and these results indicate that it is more likely to see a net cash flow from a more-developed to a less-developed province. 
We explain the effect of the distance by simple geographic proximity: nearby provinces have more interactions and migration between them.

Additionally, we also examine the overall cash flows between people of different ages and find that, generally, the net cash flow is from older to younger people. This is consistent with social norms in China, i.e.,
older people give red packets to younger people, especially to children and the unmarried. In terms of gender, we find that, in general, the direction of the net cash flow is from males to females. Comparing four types of cash flow (male to male, male to female, female to male, and female to female), we find that male to male represents the largest proportion (36\%). Details on cash flows among age and gender groups are stated in SI.

\subsection{Demographic Factors and Gift Giving}

In addition to overall cash flows, we examine the 
motivation for sending red packets by distinguishing two types of red packets and the effect of different factors 
on the likelihood of sending red packets.

\begin{figure}
\includegraphics[width=\linewidth]{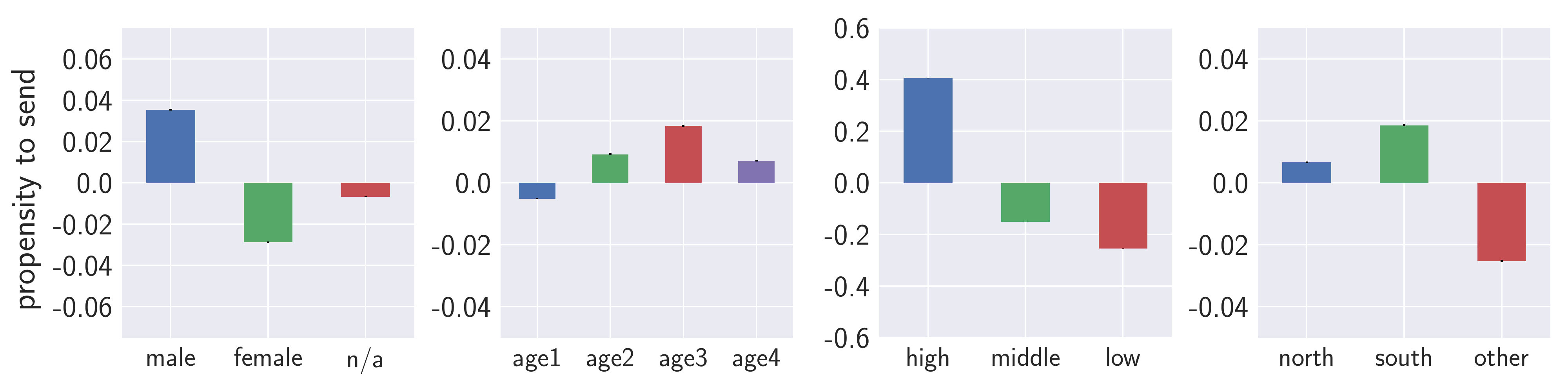}
\caption{Propensity to send given (a) gender (b) age (c) degree (d) location. We define three groups: \textit{high-d} (people with in-group degree that is no less than the second tertile), \textit{low-d} (people with in-group degree that is no more than the first tertile) and \textit{mid-d} (the rest). Group-level SEs are shown as blacks ticks on the tops/bottoms of the bars.}
\label{fig:send}
\end{figure}

\begin{figure}
\includegraphics[width=\linewidth]{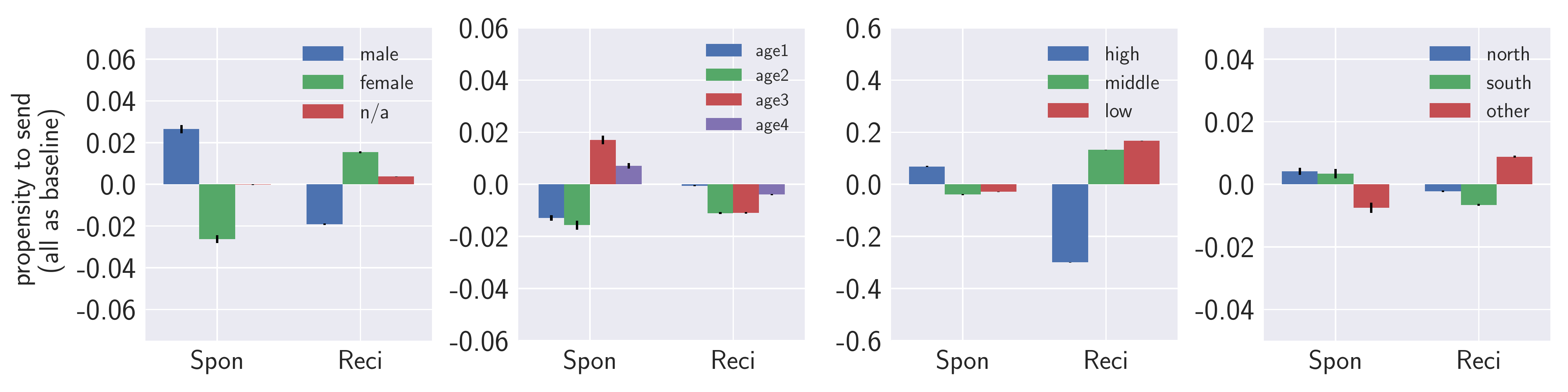}
\caption{Relative Propensity to send spontaneous/reciprocal red packets given (a) gender, (b) age, (c) degree, and (d) location. We define three groups: \textit{high-d} (people with in-group degree that is no less than the second tertile), \textit{low-d} (people with in-group degree that is no more than the first tertile) and \textit{mid-d} (the rest). Group-level SEs are shown as blacks ticks on the tops/bottoms of the bars.}
\label{fig:send_sr}
\end{figure}

Reciprocity is considered an important motive for gift giving\cite{komter1996reciprocity}. We distinguish a particular type of red packets: reciprocal red packets. If a person receives a red packet and then quickly sends a new red packet, we consider the new red packet to be reciprocal. We also define spontaneous red packets, which are initiated without the sender having recently received a red packet, and thus, these are unlikely to be triggered by reciprocity.
The formal definitions are as follows:
\begin{itemize}
\setlength\itemsep{0.1em}

\item Reciprocal red packets. We define a reciprocal red packet based on the following three criteria. (1) The time interval between this red packet and the prior packet is no more than 600 seconds; (2) before sending this red packet, the sender had received a red packet in the previous 600 seconds; (3) the two consecutive packets are not sent by the same person. In other words, if we observe someone receiving a red packet from another group member in a short period and sending a red packet to the same WeChat group, 
it is likely that she is motivated by reciprocity.

\item Spontaneous red packets. If the time interval between a red packet and the prior red packet is more than 24 hours, we consider it to be a spontaneous red packet. It is very unlikely that users send such red packets to express reciprocity, and one possible motivation for sending them is to promote group communication.

\end{itemize} 
We define \textit{propensity to send given feature $F$} (see Methods). Fig.~\ref{fig:send} presents the average propensity score for key features across all groups. 
We then define the relative propensity to send each particular type of red packets given a feature $F$. Specifically, we subtract the propensity to send any red packet from the propensity to send a spontaneous/reciprocal red packet. The results are shown in Fig.~\ref{fig:send_sr}. 
Details of definition of the variables can be found in SI.

We have the following observations based on Figures ~\ref{fig:send} and ~\ref{fig:send_sr}. Please refer to SI for the definition of related variables.

\begin{itemize}
\setlength\itemsep{0.1em}
\item Gender:  In general, men are more likely to send red packets than women. More importantly, this gender difference is stronger for {\em spontaneous red packets}. This suggests that the motivation for sending red packets
differs across genders, i.e., men are more likely to initiate sending behavior rather than reciprocate others' red packets.

\item Age:  Users between 30 and 40 years of age are the most likely to send red packets, especially spontaneous ones. Users less than 30 years of age are less likely to send red packets, and their red packets are more likely to be reciprocal. Users over 40 years of age are less active due to their underrepresentation and overall inactivity on WeChat. 
\item In-group degree: 
A group member of high degree (a proxy of in-group status) is more likely to send a red packet. However, they are less likely to send reciprocal red packets. One explanation is that  she might be the creator of a group and tend to initiate red packets more often than others. 
Also, high-status people may be concerned more with their prestige and thus are more willing to spontaneously  reward other group members instead of following others. 

\item Location. Consistent to results in cash flow, southerners are more likely to send red packets, and this holds for both two types of red packets. 
\end{itemize}

We have similar analysis on receivers, which can be found in \textit{SI}.

\subsection{Effects of Red Packets on Social Groups}

Although reciprocity is a widely accepted motivation for gift-giving, it cannot explain the motivations for all red packets. For example, sending a spontaneous red packet is unlikely to be driven by pure reciprocity.  
herefore, in this section, we attempt to measure the causal effect of spontaneous red packets on the group, as well as the red packet's sender, by applying propensity score matching to examine the possible benefit that red packets bring to groups.

Since it is difficult to conduct counter-factual analysis such as ``what if this red packet had not been sent'', we choose to match and compare the spontaneous red packets with large amounts of money compared to those with small amounts of money. Doing so allows us to determine whether a marginal increase in the money in a red packet can have significant impacts on group activities. Therefore, our treatment dummy is coded as one when the cash amount of a spontaneous red packet $\geq$ 100 RMB.\footnote{We also used other thresholds, i.e., 50 and 150 RMB, and find consistent results.}, zero otherwise. In total, we have 54,150 red packets in the treatment group and 6,133,489 in the control group. Table~\ref{tab:reg} reports the regression results. We also control for the number of splits allowed for each red packet (RPnum), and group fixed effects, including at the individual and seasonal levels.

We define a \textit{session} as a sequence of red packets in a group where the interval between two consecutive red packets is shorter than $\tau$. $\tau=30\text{min}$, and we examine the effect of spontaneous red packets on the session. 
The dependent variables are (1) the total amount of money available to following users in this session (session money); (2) the number of following users who send red packets in the session (number of following senders, or \#FS); (3) the number of following red packets in the session (\#FRP); and (4) the number of new in-group friends that the sender has within 1 day after a spontaneous red packet is sent (\#Senders' new friends).

Compared to prior studies that match on individual-level data \cite{aral2009distinguishing}, we further utilize group features, e.g., the size of the group, and seasonality features, e.g., whether the red packet is sent during the spring festival, to understand the causal effect of the amount of cash in red packets. All 54,150 treatment observations are successfully matched. The standardized differences of all covariates are below 0.1, which indicates insignificant imbalance between the treatment and control groups \cite{normand2001validating}.

\begin{table}[!htbp] \centering 
  \caption{Determinants of Group Activities} 
  
  \centering
  \scriptsize
  \label{tab:reg} 
\begin{tabular}{l D{.}{.}{-3} D{.}{.}{-3} D{.}{.}{-3} D{.}{.}{-3} } 
\setlength\tabcolsep{0.1pt}
\\[-1.8ex]\hline 
\hline \\[-1.8ex] 
 & \multicolumn{4}{c}{\textit{Dependent variable:}} \\ 
\cline{2-5} 
\\[-1.8ex] & \multicolumn{1}{c}{\#Follower} & \multicolumn{1}{c}{\#FRP} & \multicolumn{1}{c}{Session Money} & \multicolumn{1}{c}{\#Sender's New Frds} \\ 
\\[-1.8ex] & \multicolumn{1}{c}{(1)} & \multicolumn{1}{c}{(2)} & \multicolumn{1}{c}{(3)} & \multicolumn{1}{c}{(4)}\\ 
\hline \\[-1.8ex] 
 treatment (large amount) & 0.295^{***} & 0.415^{***} & 52.632^{***} & 0.009^{***} \\ 
  & (0.010) & (0.024) & (1.490) & (0.003) \\ 
  is festival & 0.185^{***} & 0.224^{***} & 6.479^{*} & -0.006 \\ 
  & (0.025) & (0.057) & (3.625) & (0.008) \\ 
  is weekday & -0.094^{***} & -0.078^{***} & -4.862^{***} & -0.002 \\ 
  & (0.011) & (0.025) & (1.602) & (0.003) \\ 
  RPnum & 0.006^{***} & 0.006^{***} & 0.218^{***} & 0.0003^{***} \\ 
  & (0.0004) & (0.001) & (0.055) & (0.0001) \\ 
  female & 0.089^{***} & 0.119^{***} & 9.683^{***} & 0.002 \\ 
  & (0.013) & (0.029) & (1.815) & (0.004) \\ 
  degree & -0.145^{***} & -0.388^{***} & -12.416^{***} & -0.039^{***} \\ 
  & (0.028) & (0.065) & (4.111) & (0.009) \\ 
  group size & 0.0004^{***} & 0.002^{***} & 0.015 & 0.001^{***} \\ 
  & (0.0001) & (0.0002) & (0.012) & (0.00002) \\ 
  & (2.349) & (5.396) & (340.545) & (0.716) \\ 
  group fixed effect & Y & Y & Y & Y \\
  individual fixed effect & Y & Y & Y & Y \\
  seasonal effect & Y & Y & Y & Y \\
 \hline \\[-1.8ex] 
Observations & \multicolumn{1}{c}{108,300} & \multicolumn{1}{c}{108,300} & \multicolumn{1}{c}{108,300} & \multicolumn{1}{c}{108,300} \\ 
R$^{2}$ & \multicolumn{1}{c}{0.036} & \multicolumn{1}{c}{0.019} & \multicolumn{1}{c}{0.016} & \multicolumn{1}{c}{0.051} \\ 
\hline 
\hline \\[-1.8ex] 
\multicolumn{5}{l}{$^{*}$p$<$0.1; $^{**}$p$<$0.05; $^{***}$p$<$0.01} \\ 
\multicolumn{5}{l}{Group/individual/seasonal effects refer to all control variables not listed above.} \\
\end{tabular} 
\end{table}

First, compared to small red packets, on average, large red packets have 0.295 more group members following and generate 0.415 more following red packets (Columns 1 and 2). Additionally, red packets sent on special days, such as festivals and weekends, generate more following users and following red packets. 
Interestingly, those sent by female users and by people with lower in-group degree also have more following users and more following red packets. 

Second, we examine the impact on the following session money (Column 3). We find that a large red packet, on average, increases 52.632 RMB in session money. This indicates that a large cash amount may significantly benefit other group members. Therefore, they might be more likely to reciprocate by sending red packets to the group. 

Finally, we analyze the number of new friends that a group member garners by sending a large amount of red packets. On average, a large amount of red packets attracts 0.009 new friends to the sender. This indicates that red packets can help senders to attract new friends.

\section{Discussion}

In summary, we have presented the macro- and micro-level patterns in red packet sending and discussed the motivations for and benefits of sending red packets.

Studying the interprovincial cash flows improves our understanding of human mobility. 
The patterns that we found are consistent with a gravity model: the interactions between two areas, for example in the form of immigration and trade, are proportional to the ``masses'' in these two areas (population, economic development) and inversely proportional to the distance between them\cite{lewer2008gravity}; similarly, cash flows between provinces are dependent on both living standards (PPP) and distance. 
We conjecture that the trends could be driven by immigration: people move to nearby and more developed provinces and maintain their social bonds with families and friends in their hometowns.

Our micro-level results add new dimensions to our understanding of gift giving. 
For example, we find that men are generally more inclined to send red packets and that red packets from women are more likely to be reciprocal than spontaneous. 
Most existing studies on gender differences in gift exchange, especially in charitable giving, state that women are more likely to donate to more charities \cite{kottasz2004differences,andreoni2003charitable,mesch2011gender}. However, 
in the context of group red packets, men are more likely to send red packets. Rather than pure altruism, we conjecture that income inequality and a desire for prestige account for this gender difference. Although it has decreased, the gender income gap persists in China \cite{xiu2013gender}, which limits the budget available to females to reward others. Another account is the desire for prestige, or ``mianzi'' in Chinese culture. Findings indicate that men are more inclined to seek status than women \cite{von2010men}, and sending red packets, especially spontaneous red packets, provides a way to demonstrate one's generosity and wealth.

The results of differences in reciprocal and spontaneous red packets and casual effects of spontaneous red packets explain the motivations of sending red packets, 
and furthermore, reasons why online red packets has spread rapidly in China. 
We find that the increased rewards (amount of money) spontaneous red packets of increased rewards (amount of money) to a group can significant stimulate other group members to send follow-up red packets. 
Ultimately, spontaneous red packets generate rewards for group that exceed the packets' initial values. 
The explanation for this increase is likely reciprocity, that is, others have received money and feel obliged to send money in turn. 
Moreover, motivations of the spontaneous senders who tend to be senior or  people with high in-group degree can be explained by the aforementioned benefits of sending red packets.

However, our work does suffer from limitations. 
First, despite the large sample size, our dataset is unable to represent the entire population of China. Although 65\% people in China use WeChat, users who are active in red packets are biased toward young and middle-aged people. Moreover, we filtered out inactive groups and suspiciously highly active groups, which may also have influenced our results. 
Second, gender, age, location, and status differences can be confounded by unobserved factors such as income. However, because of privacy issues, we were unable to collect such information. Therefore, we are not arguing for causal effects of, for example, gender or age. 
Third, although we explored the motivations for sending red packets by examining the causal effects of spontaneous red packets, we do not have a comprehensive understanding of the true motivations of such sending behavior, although we propose that reciprocity and the beneficial effects on group activities could be two main reasons.

There are many promising directions for future research. For example, if we were to collect data from all message groups to which an individual belongs, we could explore her strategies for allocating money across the groups. In addition, we could also study how red packets are spread across all WeChat groups and which groups are responsible for the rapid adoption of red packets.

\section{Methods}
\subsection{Data Description}
We first randomly sample one million WeChat groups in which at least one group red packet was sent between October 1, 2015, and February 29, 2016.
To avoid the problem of data sparsity, we exclude
groups with fewer than $\eta$ red packets. We further exclude extremely active groups with more than $\tau$ red packets as a group that have extremely large amounts of money in the red packets, as these might be used for online gambling, and incorporating these groups might significantly bias our results. Specifically, we empirically set $\eta = 3 \times m$ and $\tau=50 \times m$, where $m$ is the group size.
In total, this selection process leaves us with 367,361 groups with 7,816,214 group members (7,380,110 unique users).

We combine three datasets for our statistical analyses, including (1) the characteristics of 367,361 WeChat groups, e.g., the number of members, total number of red packets and the total value of the red packets; (2) the characteristics of the 7,380,110 unique users in these WeChat groups, e.g., demographic variables, the number of friends and the number of WeChats that they join; and (3) propensities for red packets, such as the cash amount and the number of recipients of a particular red packet. In total, 61,501,862 red packets were sent in these WeChat groups, generating 218,301,860 recipients. 


During this five-month period, the average number of red packets sent in a group is 172.88 and the average amount of money exchanged in a group is 995.78 RMB (approximately 146 USD).  All details on the data and summary statistics of the variables can be found in the Data section of the \textit{SI}.

\subsection{Propensity to Send}
We define a group $g \in G$, where $R_g$ represents number of red packets sent in group $g$. 
Among all group members (potential senders), there are $N_F$ group members with the feature in question and $N_{\bar{F}}$ without it. 
Among all red packets in that group, we can compute the frequency of red packets sent by people with feature $F$, $T_F$, and without feature $F$, $T_{\bar{F}}$. Then, for each group, we compute $\frac{T_F}{T_F+T_{\bar{F}}} - \frac{N_F}{N_F+N_{\bar{F}}}$ as the propensity to send given feature $F$ in this group. 

\subsection{Covariates in Matching}
To eliminate the influence of other factors, we extract various features for matching. For the group features (group-level fixed effect), we have the size of the group, density of the group network at the time red packet is sent, gender ratio, age entropy, and province entropy. For the sender features (individual-level fixed effect), we have the gender, age, location of the city (latitude and longitude), the number of WeChat friends the individual has (activity\_1), the number of WeChat groups to which the individual belongs (activity\_2), and the degree ratio ($\text{degree}/(\text{member count} - 1)$). For the seasonal features, we have whether it is during a festival (such as the spring festival period, New Year's Day, and Christmas), whether it is weekday, the period of a given day (we divide one day into six equal periods, as an integer; empirically, a higher number indicates a higher probability to send more and larger red packets), and the UNIX time stamp of the sent red packet (how many seconds from  January 1, 1970, sendtime). Additionally, we have the number of parts into which the red packet was divided (RPnum). Imposing such controls can help to exclude the trivial statements such as  ``during festivals, people are more likely to send big red packets when people are also very active on WeChat to send red packets''.

\bibliographystyle{plain}
\bibliography{main}

\clearpage
\begin{appendices}
\section{Background}
Comparable to the popularity of non-cash gift-giving in Western culture, Asians have a long history of giving cash to one another to express their gratitude and deepen their relationship. 
Moreover, they typically place the cash inside a packet, the color of which differs across cultures. 
For example, the packets are red in China and Singapore and 
white in Japan.\footnote{We use the term ``red packet''  if doing so generates no ambiguity.} 

These red packets are among the most important components of festivals, for example, the Chinese lunar new year and important events such as wedding and birthdays. 
They are the most frequently used tool for gift exchange between friends
and an effective mechanism to maintain the close relationships with family members.\footnote{They have also been criticized for being instruments of bribery and corruption. This is beyond the scope of this work and will be discussed elsewhere.}

\begin{figure}
\centering
\begin{tabular}{@{}cccc@{}}
\includegraphics[width=0.22\textwidth]{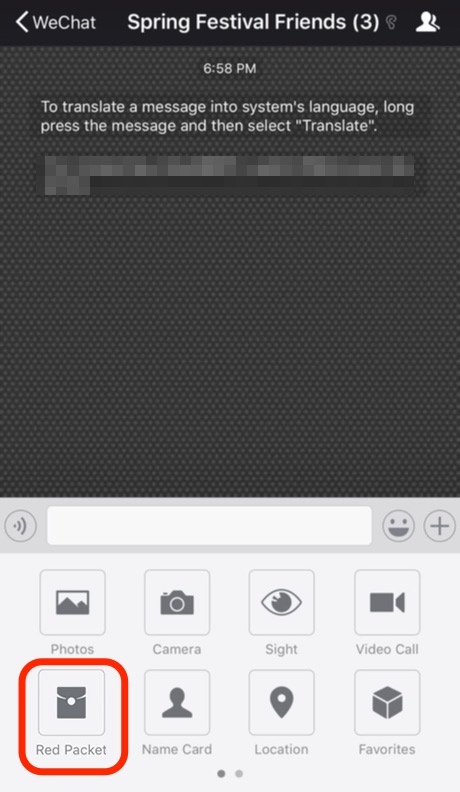}
&
\includegraphics[width=0.22\textwidth]{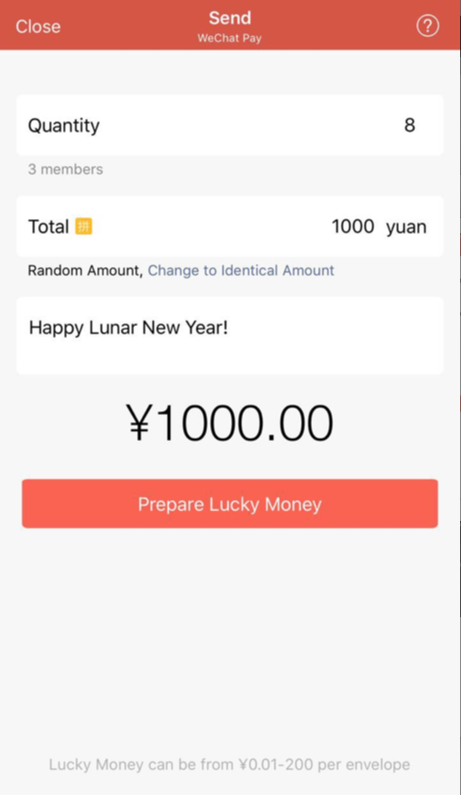}
&
\includegraphics[width=0.218\textwidth]{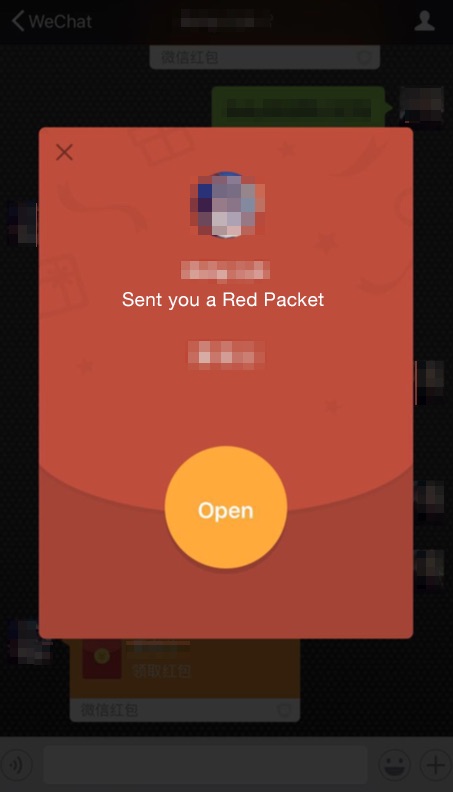}
&
\includegraphics[width=0.219\textwidth]{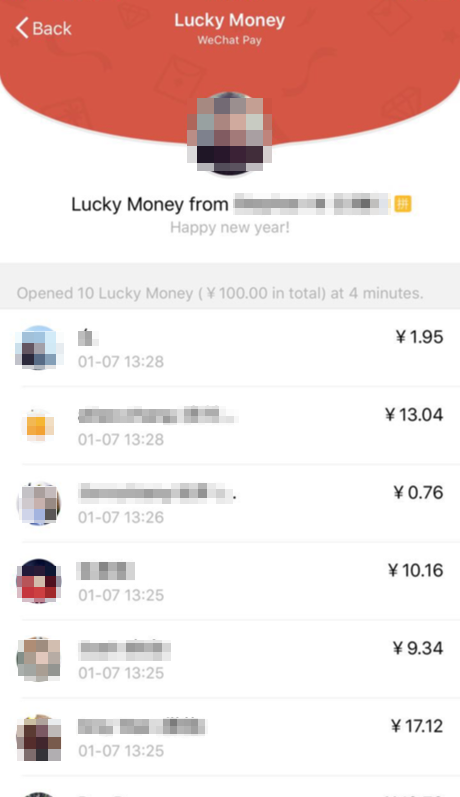}\\
(a)  & (b) & (c) & (d)
\end{tabular}
\caption{An example of sending and receiving group a red packet. Four steps are listed. The first two are seen by the sender, and the following two are seen by group members who successfully receive the red packets.
} 
\label{fig:illustration}
\end{figure}

Because of its lack of geographic constraints, WeChat significantly speeds the flow of red packets among individuals. Moreover, the popularity of WeChat \textit{groups} results in an extremely interesting change --- the cash flow is extended from one-to-one exchange to a one-to-many case, which is called the ``Group Red Packet''. Specifically, a group user can send a red packet to a WeChat group. She can either choose an identical amount or a random amount for each group member. For identical packets, she needs to specify each packet's monetary value, while for random red packets, she needs to specify the total amount. Additionally, she needs to specify the number of recipients (``quantity"), which can be fewer than the number of group members. Therefore, the lottery nature of random red packets and the possibility of not receiving a packet generate the entertainment aspect of red packets, which has made them quite popular on WeChat.

Figure~\ref{fig:illustration} presents an example of group red packets with a random amount. As shown in panel (a), a group member clicks the ``red packet'' button on the left corner of the screen. Then, a window in panel (b) pops out. Conditional on the user's choice of random amount, she determines the quantity of red packets and the total amount. She can also leave a message that will be shown at the top of the red packets, e.g., ``Happy Lunar New Year!''. Finally, she clicks ``prepare lucky money'', and the red packet is sent to the group. Thereafter, everyone in this group will be notified of this red packet (panel c) and can choose to click the ``open" button to determine the (random) amount they will receive.
Finally, all group members can determine the amount that all other members received. The person who receives the highest amount will be tagged with a star icon with the legend ``Luckiest Draw''. \footnote{See 
\url{https://goo.gl/eUjNTy} for a detailed description.}

\section{Data Description}

There are 7,380,110 unique users in our 367,361 groups, and a user may belong to multiple groups. 
We retrieve each user's demographic information, e.g., gender, age, location, the number of  friends, and the number of groups that she joins. Since these information is updated every month, we use the information retrieved in February 2016 for analyses. 
Moreover, we identify the friendship between users in the dataset. In total, there are 292,565,291 friend links in our dataset and we also accessed the date for two users become WeChat friends. 

For each group $g_i$, we retrieve group member $g_{ij}$'s frequency of sending (receiving) red packets, the cash amount of each red packet, the number of messages that she posts in group $i$, and the time stamp that she joined group $i$. Table~\ref{tab:group} reports summary statistics for groups.

As the individual level, the relevant variables are as follows. Table~\ref{tab:attribute} reports summary statistics for individual users.

\begin{itemize}
\item \textbf{Gender}:Users report their gender. In total, 3,893,537 users are male, 3,438,393 are female, and the remainder have no data. 
\item \textbf{Region and Location}. Since most WeChat users live in Mainland China, we focus on 31 provinces (or municipalities/autonomous regions) in Mainland China, excluding Hong Kong,  Macau and Taiwan. We classify the cities where a user resides in into seven categories\footnote{We refer to Wikipedia: \url{https://en.wikipedia.org/wiki/List_of_regions_of_the_People\%27s_Republic_of_China}}. (1) North contains Beijing, Tianjin, Hebei, Shanxi, and Inner Mongolia; (2) Northeast contains Liaoning, Jilin, and Heilongjiang; (3) East contains Shanghai, Jiangsu, Zhejiang, Anhui, Fujian, Jiangxi, and Shandong; (4) Center contains Henan, Hubei, and Hunan; (5) Southwest contains Chongqing, Sichuan, Guizhou, Yunnan, and Tibet; 
(6) Northwest contains Shaanxi, Gansu, Qinghai, Ningxia, and Xinjiang;
and (7) South contains Guangdong, Guangxi, and Hainan. Although it is imprecise, we further cluster the North, Northwest and Northeast regions as ``North'' and the rest as ``South''.
\item \textbf{Age}. WeChat reports that they collect and infer users' age with over 70\% accuracy. We classify people into six age groups: $[10, 20)$, $[20, 30)$, $[30, 40)$, $[40, 50)$ and $[50, 60)$, and $[60, 70)$, which we name \textit{age1}, \textit{age2}, \textit{age3}, \textit{age4}, \textit{age5}, \textit{age6}, respectively. As shown in Table~\ref{tab:attribute}, the sizes of the age groups are unbalanced. 
\item \textbf{Degree Ratio}. Group members in a WeChat group ($G_{i}$) can be either friends or strangers on WeChat. Therefore, a WeChat group is an undirected network, where each user $j$ is a node and an edge represents a ``friend'' relationship between two group members. The degree of a group member $j$ in group $i$, $d_{ij}$, is the number of friends that she has in this group, and we define her degree ratio in a group $r_{ij}$ as $\frac{d_{ij}}{\# \mbox{group member} - 1}$. $r_{ij} \in [0, 1]$. The higher $r_{ij}$ is, the more closely a user $j$ is attached to group $i$. 
\end{itemize}

\begin{table*}[t]
	\centering
	\caption{Statistics of Red Packet Groups.}
	\label{tab:group}
\begin{threeparttable}
		\begin{tabular}{c|cccccc}
			\hline \hline
			Statistics       & Mean   & Min  & 25\%   & 50\%   & 75\%    & Max       \\ \hline
			\#member             & 13.58  & 3    & 5      & 10     & 17      & 500       \\
			Density\tnote{a}          & 0.55   & 0    & 0.31   & 0.5    & 0.775   & 1         \\
			Total RP money      & 995.78 & 0.09 & 148.31 & 441.81 & 1065.11 & 492,969.24 \\
			Total RP count      & 172.88 & 9    & 46     & 101    & 184     & 16,795     \\
			RP Money per capita & 96.82  & 0.03 & 15.76  & 40.13  & 101.94  & 18,600     \\
			RP Count per capita & 12.98  & 3    & 5.24   & 9.31   & 17.33   & 50   \\ 
			Female ratio & 0.53 & 0 & 0.4 & 0.52 & 0.67 & 1.0 \\ 
			Age group entropy\tnote{b} & 0.75  & 0.0    & 0.29   & 0.83   & 1.07   & 1.94   \\ 
			Location region entropy & 0.40  &  0.0   & 0.0   & 0.04   &  0.83  & 1.97   \\ 
			\hline \hline
		\end{tabular}
		\begin{tablenotes}
			\item[a] Density measures the connectivity within a group. 
			It is defined as $\frac{\text{\# pairs of friends}}{\text{(\#member)(\#meber-1)}/2}$
			\item[b] Entropy is calculated as $\sum_i -p_i \log_2(p_i)$, where $p_i$ is the proportion of each age/province group.
		\end{tablenotes}
	\end{threeparttable}
\end{table*}

\begin{table}[t]
\centering
\caption{Statistics of WeChat Users}
\label{tab:attribute}
\begin{tabular}{lll}
\hline\hline
\textbf{Demographic features} & \textbf{Count}     & \textbf{Proportion} \\ \hline
\textit{Gender}               &           &            \\
Male                 & 3,893,537 & 52.76\%    \\
Female               & 3,438,293 & 46.59\%    \\
Unreported           & 48,280    & 0.65\%     \\
\textit{Location}             &           &            \\
North                & 795,113   & 10.77\%    \\
Northeast            & 465,941   & 6.31\%     \\
Northwest            & 433,681   & 5.88\%     \\
East                 & 1,462,468 & 19.82\%    \\
Center               & 937,148   & 12.70\%    \\
Southwest            & 593,279   & 8.04\%     \\
South                & 894,397   & 12.12\%    \\
Other/Unreported     & 1,798,083 & 24.36\%    \\
\textit{Age}                  &           &            \\
{[}10, 20)           & 747,226   & 10.12\%    \\
{[}20, 30)           & 3,886,688 & 52.66\%    \\
{[}30, 40)           & 1,963,130 & 26.60\%    \\
{[}40, 50)           & 488,033   & 6.61\%     \\
{[}50, 60)           & 63,410    & 0.86\%     \\
{[}60, 70)           & 4,067     & 0.06\%     \\
Other/Unreported     & 227,556   & 3.08\%     \\ \hline\hline
\end{tabular}
\end{table}

In total, 61,501,862 red packets were sent in all groups in our dataset, and red packets were received 218,301,860 times. 
The average number of red packets sent in each group is 172.88, and the average amount of money is 995.78 RMB (146 USD). Table~\ref{tab:rp} reports further statistics about the red packets.

\begin{table*}[t]
	\centering
	\caption{Statistics of Red Packets}
	\label{tab:rp}
	\begin{threeparttable}
		\begin{tabular}{c|cccccc}
			\hline\hline
			Statistics       & Mean   & Min  & 25\%   & 50\%   & 75\%    & Max       \\ \hline
			Red packet money              & 5.55  & 0.01    & 1      & 2     & 5      & 10000       \\
			Red packet parts          &  5.06  & 1    & 3   & 5    & 5   & 100         \\
			Interval\tnote{a}\  & 25,458.09 & 0 & 34 & 89 & 511 & 12,475,671    \\
			Finished time\tnote{b} & 27.92 & 1 & 6 & 8 & 11 & 86,400 \\ \hline\hline
		\end{tabular}
		\begin{tablenotes}
			\item[a] The time between the current and preceding red packet. Unit is seconds. We do not consider the first red packets in the groups that we can observe in the dataset.
			\item[b] The interval between the time when the red packet is sent and the time when the previous packet was received. Red packets that were not received by anyone are not taken into account.
		\end{tablenotes}
	\end{threeparttable}
\end{table*}

\section{Cash Flows Between Demographic groups}

Figure~\ref{fig:demoflow}(a) depicts the cash flows between gender pairs, i.e., MM (male to male), MF (male to female), FM (female to male) and FF (female to female). 
The direction of the arrow represents the direction of a cash flow, and the thickness refers to the percentage of each pair in the total cash flow between two arbitrary individuals. 
Specifically, 36.1\% of total cash flows is within men, followed by MF (24.8\%), FF(19.5\%) and FM (19.3\%).
In total, the high percentage of within-gender cash flows might be driven by homophily---people tend to make friends who are similar to them, as demonstrated by Figure~\ref{fig:recv}.
The percentage of cash flows is determined by who is more likely to send a red packet, who is more likely to participate in receiving a red packet, and the gender composition of a group. If we want to construct a model of the gender composition of a cash flow, e.g., how likely a red packet between a male and female is, we need to consider these three factors.

Figure~\ref{fig:demoflow}(b) illustrates the net cash flows between different age groups. Since the age distribution on WeChat is extremely unbalanced, e.g., users in [20, 30) represent approximately 50\% of users, we utilize balance ratio to measure the cash flows among age groups.
The balance ratio is defined as equation~\ref{equ:br}, ranging from $-1$ to $1$, and a positive balance ratio from province (gender / age) A to B implies that users from A are more likely to send red packets to those from B. We utilize the ratio to measure the significance of the cash flows among different demographic groups.
\begin{equation}
\mbox{Balance Ratio} (A\rightarrow B) =  \frac{ 2 \times \mbox{Cash} (A\rightarrow B)}{\mbox{Cash} (A\rightarrow B) + \mbox{Cash} (B\rightarrow A) } - 1
\label{equ:br}
\end{equation}

For example, the balance ratio from group 4, aged [40, 50), to group 1, [10, 20), is 0.41, indicating that the balance ratio from group 1 to group 4 is -0.41. The balance ratio from group 4 to other groups is always positive, while the balance ratio from group 1 to other groups is always negative (the arrow in Figure 5 ends at group 1). This is consistent with social norms in China, whereby older people give red packets to younger people, especially to children and the unmarried (citation). The only exception is the net cash flow from group 4 to group 5, which might be driven by the fact that those aged 40 to 50 are the breadwinners in a family and have to give red packets to both older and younger people.

\begin{figure}
\end{figure}

\begin{figure*}
\centering
\begin{tabular}{cc}
\includegraphics[width=0.48\textwidth, trim={0 0 0 0},clip]{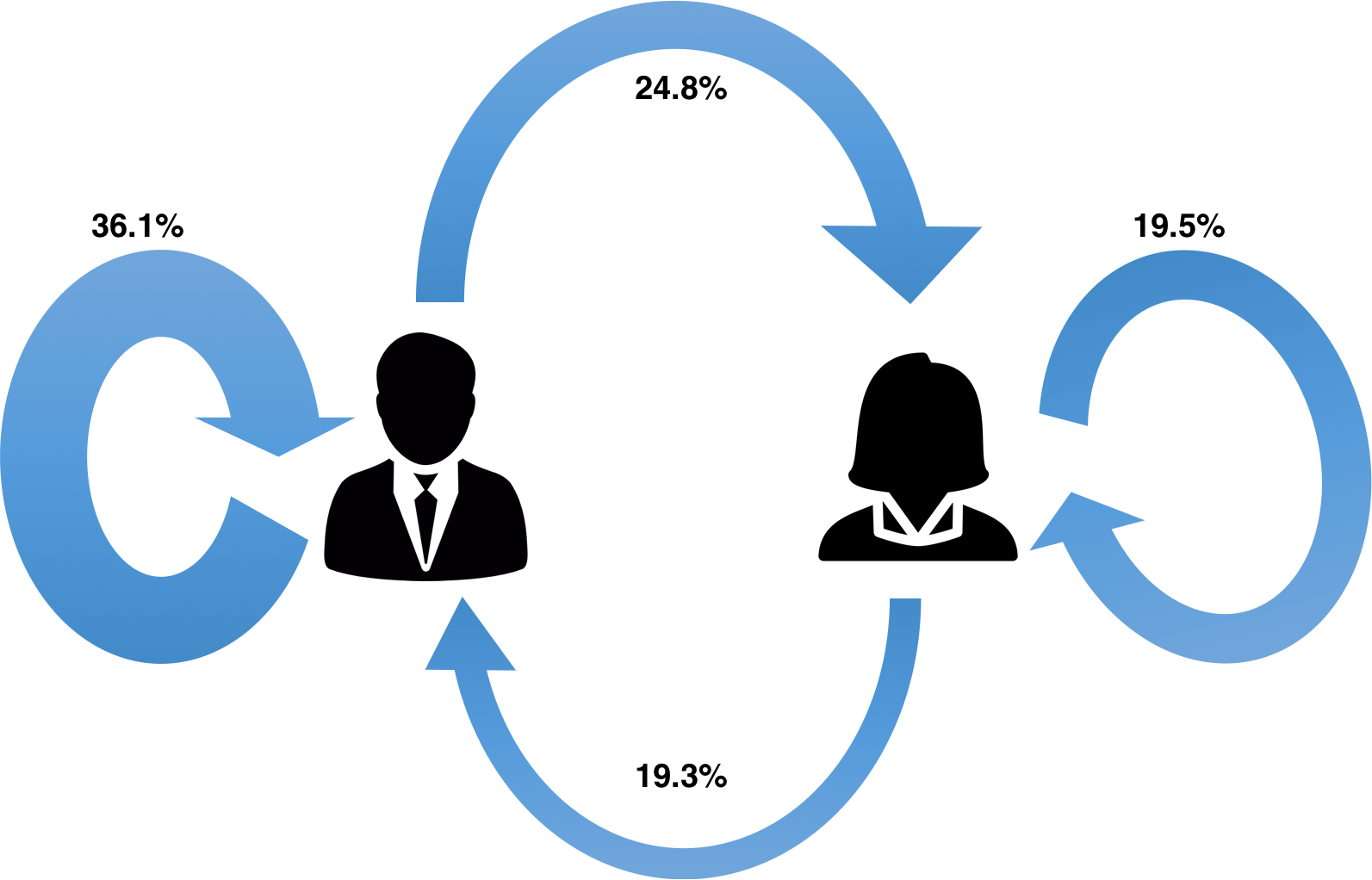} &
\includegraphics[width=0.48\textwidth, trim={20cm 0 20cm 0},clip]{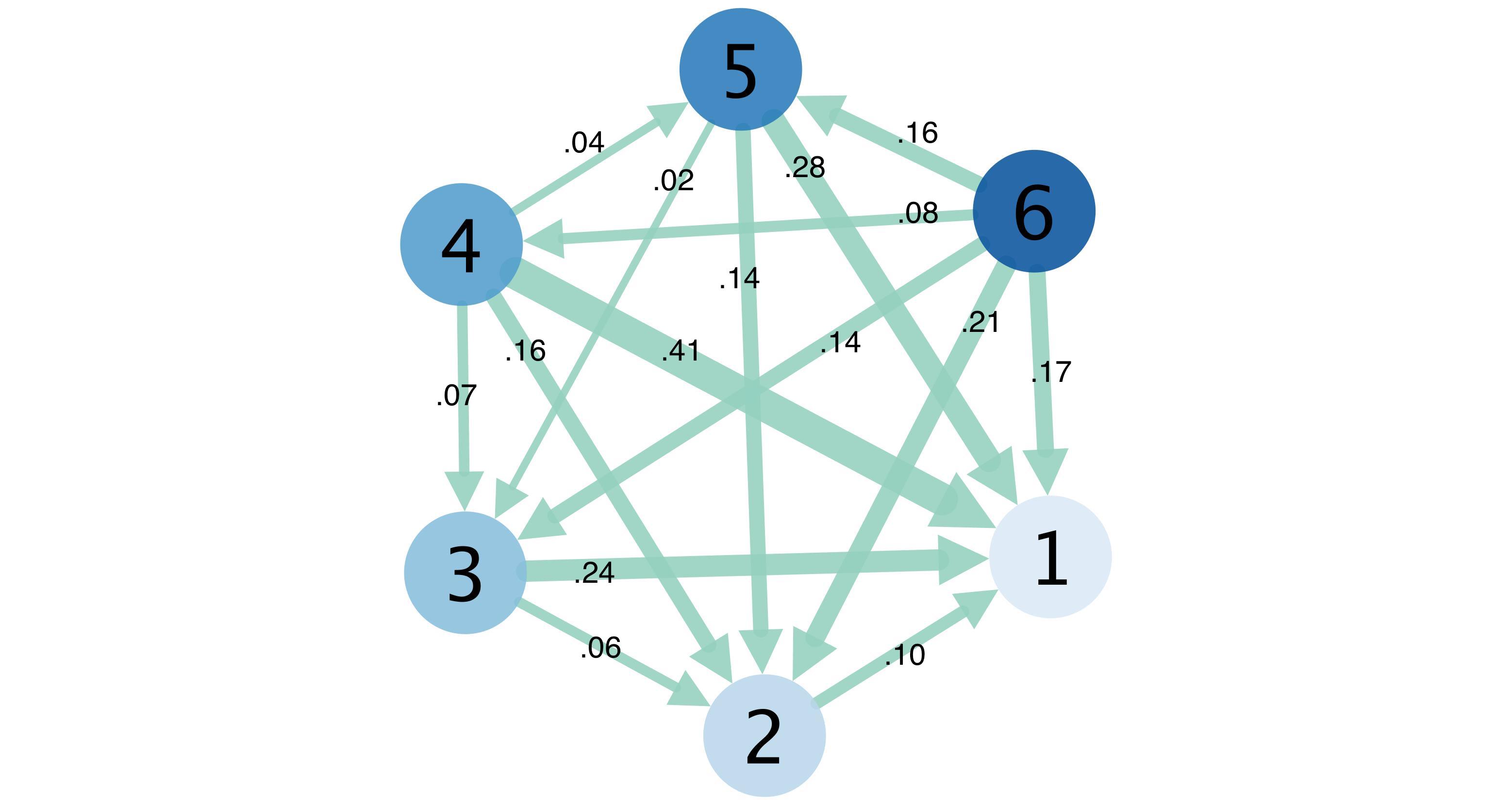} \\
(a) & (b) 
\end{tabular}
\caption{Cash flows across gender (a) and age (b) groups. The thickness of a line represents the weight of the cash flow. In (a), the curves indicate that the cash flow is from the group at the origin of an arrow to the group at the arrowhead. The percentages on the curves indicate the proportion of that cash flow in the total cash flow (we filter out cases in which people receive red packets from themselves). In (b), the number represents an age group. * represents the age group [*0, *9]. The numbers on the edges represent the net cash flow between two age groups, normalized by the maximal value, that is, from 40 to 10.}
\label{fig:demoflow}
\end{figure*}

\begin{figure}
\centering
\includegraphics[width=0.48\textwidth]{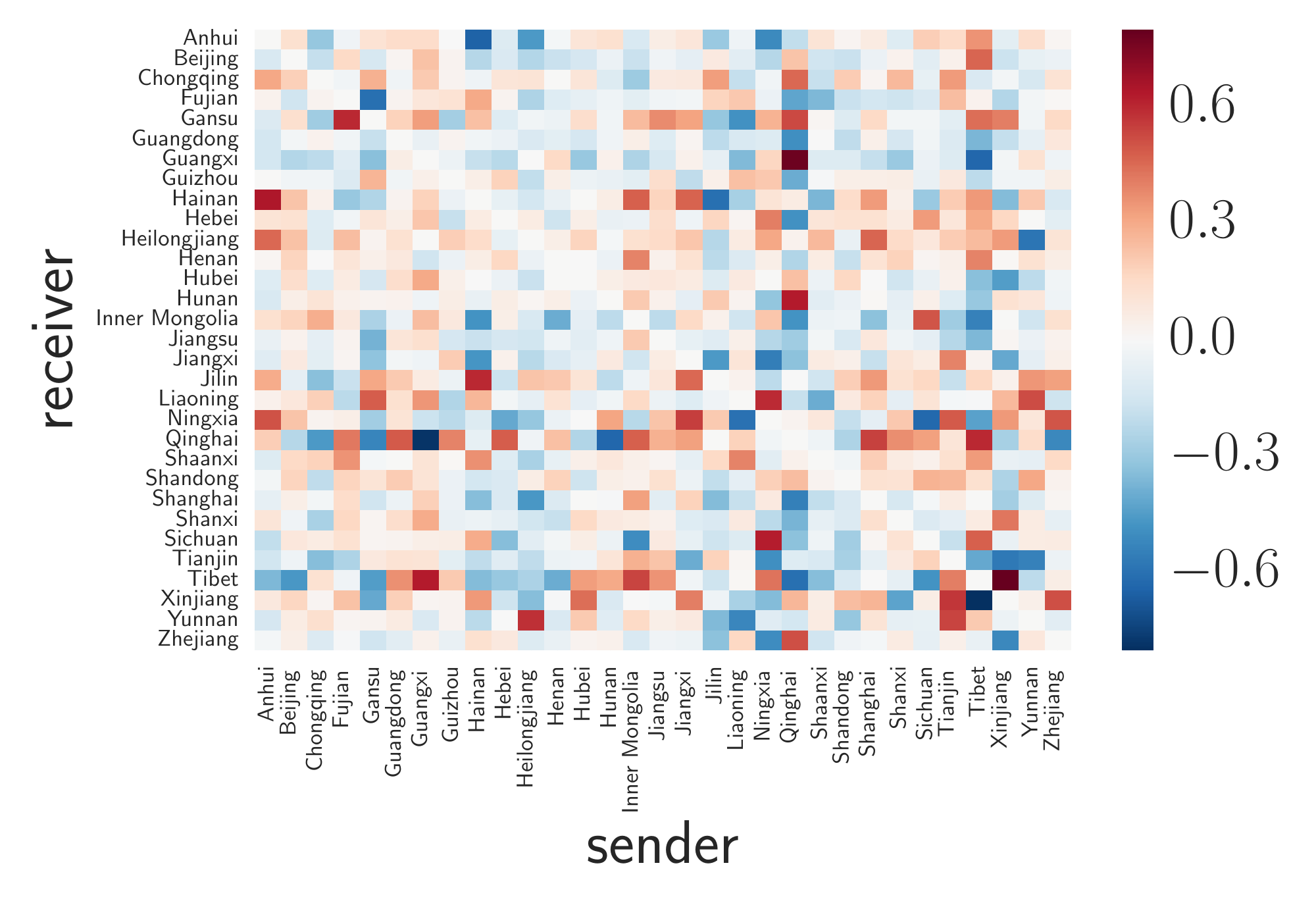}~
\includegraphics[width=0.48\textwidth, trim={4cm 2cm 10cm 4cm},clip]{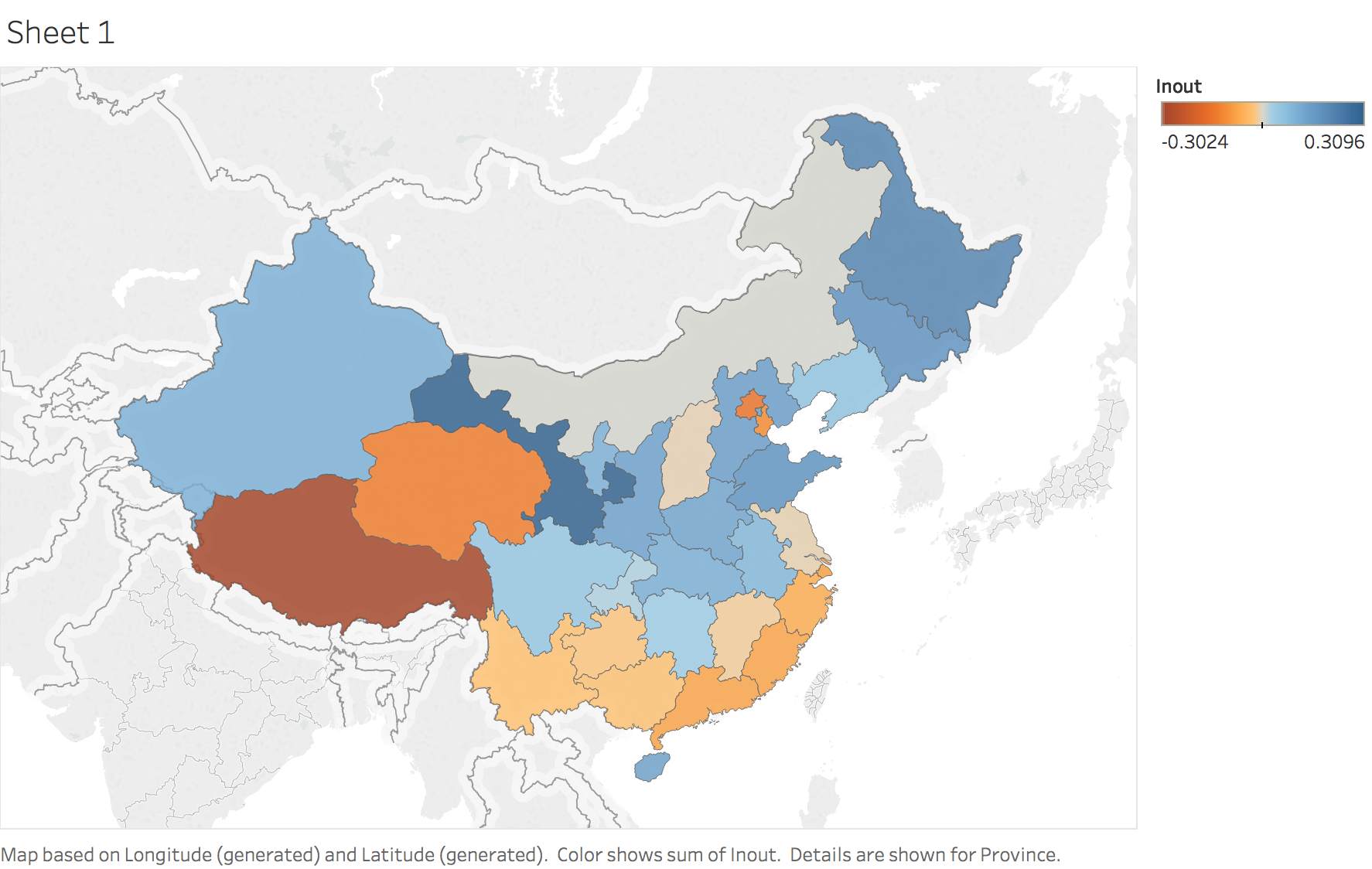}
\caption{Subfigure(a) presents the heatmap of balance ratio between provinces. Red indicates a balance ratio greater than 0, blue represents one less than 0, and gray represents a ratio of 0. Subfigure(b) presents the balance ratio between a given province and all other provinces. Red indicates a positive and blue indicates a negative ratio. The darkness of red and blue represents the absolute value of the balance ratio. }
\label{fig:province}
\end{figure}

The detailed net cash flows among location groups are shown in Figure~\ref{fig:province}. We conclude that southern and southwestern provinces, along with Beijing and Tianjin, send more money to other provinces than they receive, while the rest are net ``gainers'' from red packets. The reason that Xinjiang, Qinghai and Gansu have high absolute values of the balance ratio is partly due to the low number of red packets in the sample for most western provinces. 

For the regression analysis, we define the dependent variable as the net cash flow in the dataset between a pair of provinces (A and B), i.e.,  $|\text{Cash}(\text{A} \rightarrow \text{B}) - \text{Cash}(\text{B} \rightarrow \text{A})|$.  To remove redundant observations, we require that A has a nonnegative net cash flow to B.
We use the difference in GDP ($\text{GDP(A)} - \text{GDP(B)}$) and GDP per capita (PPP, $\text{PPP(A)} - \text{PPP(B)}$) and the distance between the provincial capitals. Table~\ref{tab:net_province} presents the regression results. We find that GDP is not a significant predictor, while PPP and distance are positively and negatively related to the net cash flow, respectively.

\begin{table*}[h] 
\centering 
  \caption{Regression Results for Interprovincial Cash Flows} 
  \label{tab:net_province} 
\begin{tabular}{@{\extracolsep{5pt}}lD{.}{.}{-3} D{.}{.}{-3} D{.}{.}{-3} D{.}{.}{-3} } 
\\[-1.8ex]\hline 
\hline \\[-1.8ex] 
 & \multicolumn{4}{c}{\textit{Dependent variable: Net Cash Flow}} \\ 
\cline{2-5} 
\\[-1.8ex] & \multicolumn{1}{c}{(1)} & \multicolumn{1}{c}{(2)} & \multicolumn{1}{c}{(3)} & \multicolumn{1}{c}{(4)}\\ 
\hline \\[-1.8ex] 
GDP      & 0.045    & -0.007   &           & 0.010    \\
         & (0.031)     & (0.034)     &           & (0.033)    \\
PPP      &           & 9.995^{***}    &           & 9.020^{***}    \\
         &           & (2.713)     &           & (2.642)    \\
Distance &           &           & -0.584^{***}   & -0.574^{***}   \\
         &           &           & (0.108)     & (0.107)    \\
Constant & 1073.916^{***} & 1020.960^{***} & 1887.943^{***} & 1810.117^{***} \\
         & (80.712)    & (80.926)    & (167.357)   & (166.762) \\
 \hline \\[-1.8ex] 
Observations & \multicolumn{1}{c}{465} & \multicolumn{1}{c}{465} & \multicolumn{1}{c}{465} & \multicolumn{1}{c}{465} \\ 
R$^{2}$ & \multicolumn{1}{c}{0.004} & \multicolumn{1}{c}{0.033} & \multicolumn{1}{c}{0.059} & \multicolumn{1}{c}{0.090} \\ 
Adjusted R$^{2}$ & \multicolumn{1}{c}{0.002} & \multicolumn{1}{c}{0.029} & \multicolumn{1}{c}{0.057} & \multicolumn{1}{c}{0.084} \\ 
\hline 
\hline \\[-1.8ex] 
\textit{Note:}  & \multicolumn{4}{r}{$^{*}$p$<$0.1; $^{**}$p$<$0.05; $^{***}$p$<$0.01} \\ 
\end{tabular} 
\end{table*} 

\section{Receiving patterns}
Users are free to choose whether to receive red packets in WeChat groups.
In some cases, users may race to collect red packets to secure one of a limited number of packets.
In this subsection, we focus on the association between the propensity to receive a gift and other variables, including demographic variables (gender, age and location), network structural variables (degree and the number of friends) and their interaction with group variables and sender variables.

In particular, we study the effects of individual variables on the propensity to receive a red packet, conditional on the characteristics of the senders. 
Similar to the propensity to send, we define the \textit{propensity to receive} of a characteristic (or demographic group).
We first define the revealed \textit{propensity to receive} a red packet as the difference between the proportion of this demographic group among receivers of red packets and the proportion of this demographic group in the group chat. 
We take the average of the propensity of red packets for each group chat and, furthermore, take the average of each group. If the aggregated propensity is 0, then we believe this demographic group has the same propensity to receive as the average WeChat user. A positive sign indicates that this demographic group is more likely to receive red packets.
Note that although WeChat allows the red packet senders to receive packets from themselves, we do not regard them as receivers or candidate receivers.

\begin{figure}
\centering
\begin{tabular}{ccc}
\includegraphics[width=0.33\linewidth]{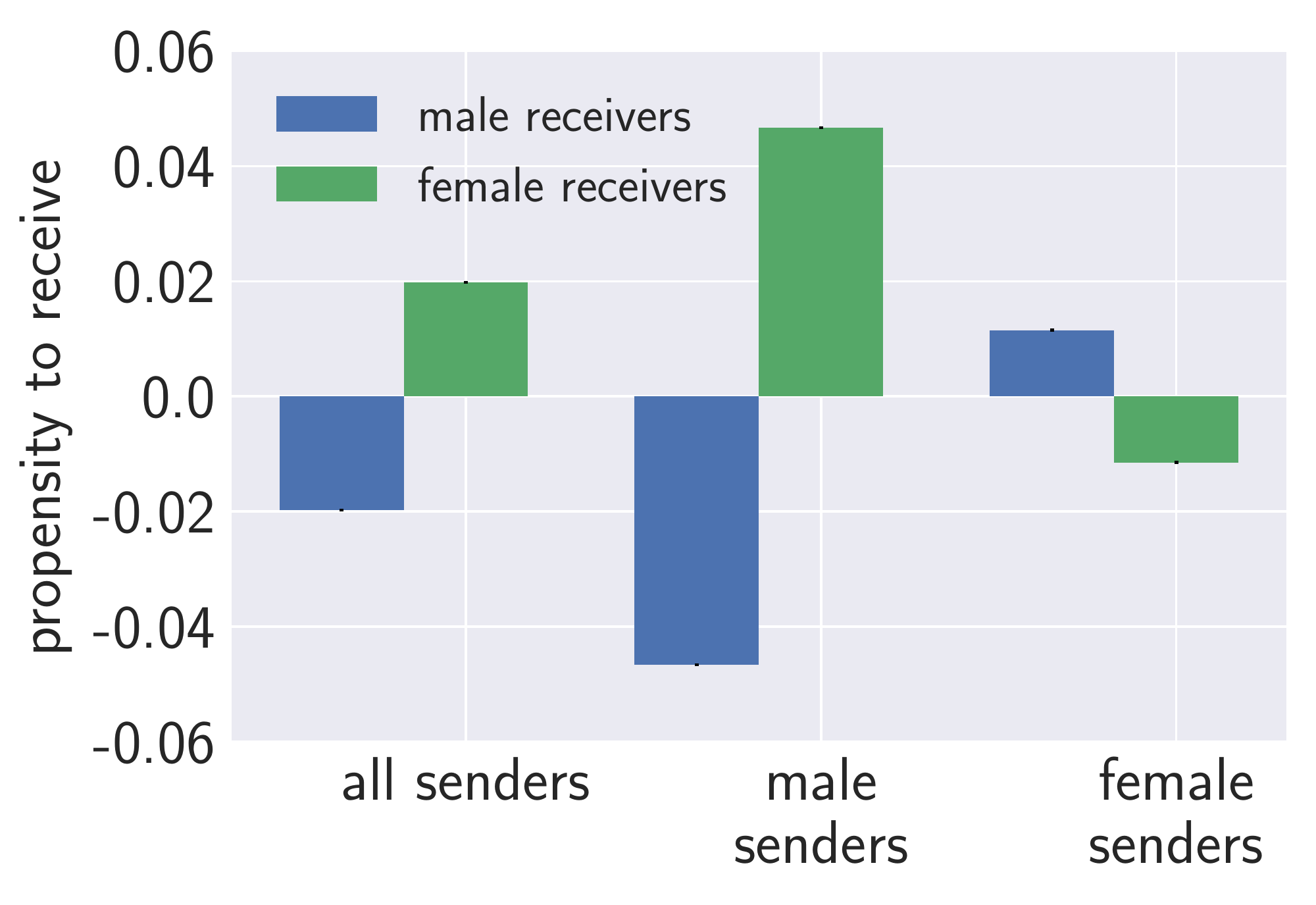} &
\includegraphics[width=0.33\linewidth]{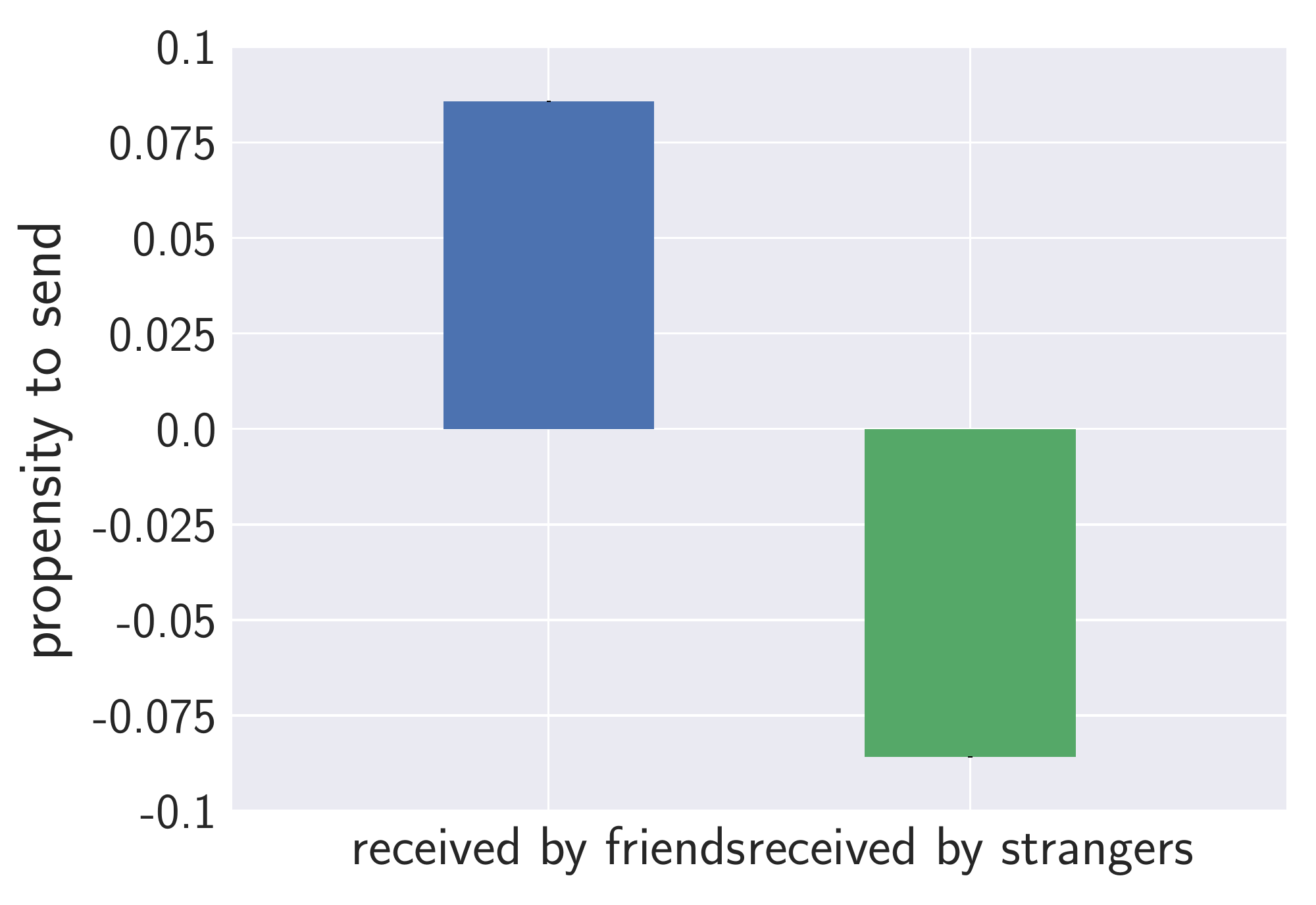} &
\includegraphics[width=0.33\linewidth]{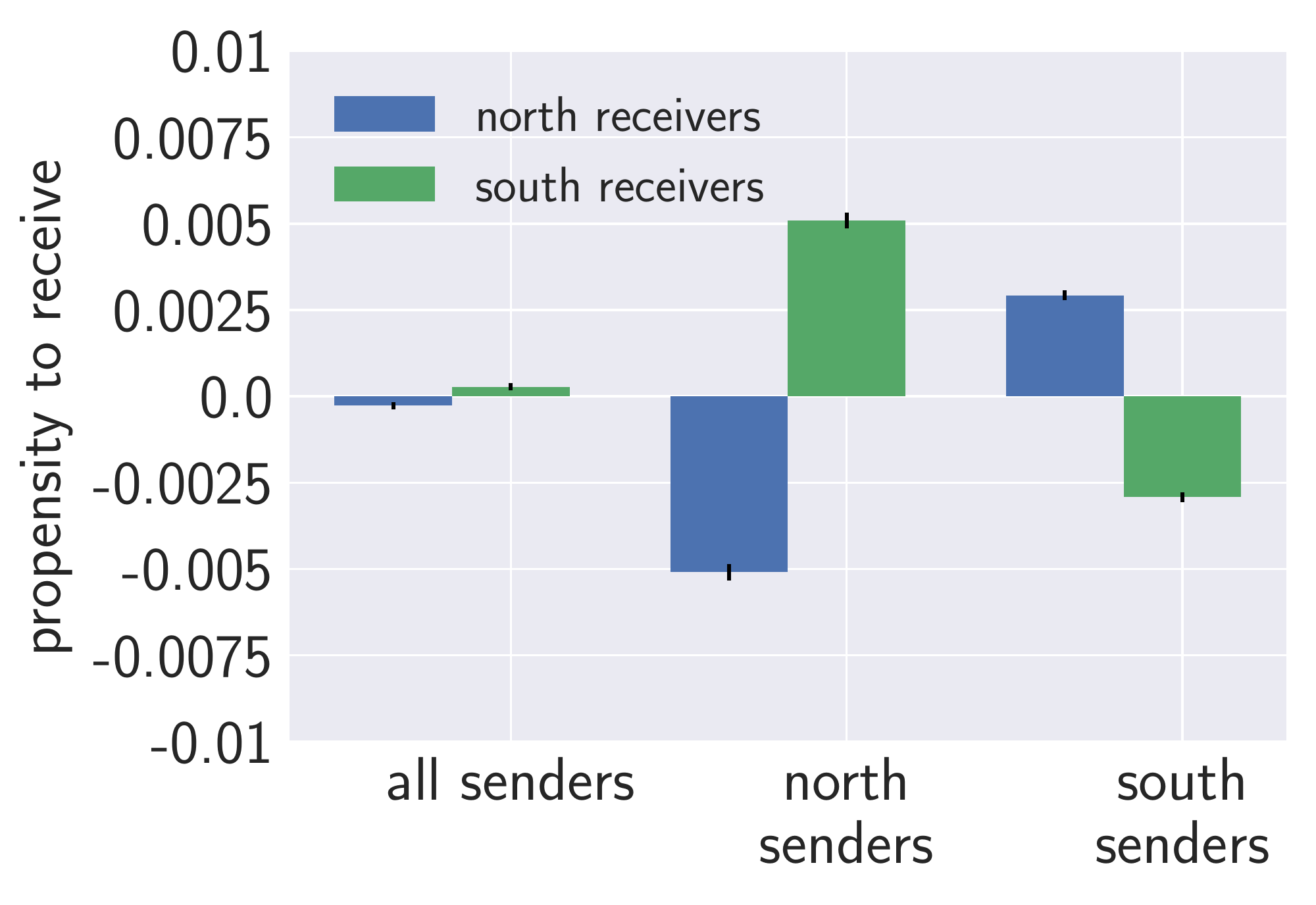} \\
(a) & (e) & (d) 
\end{tabular}
\begin{tabular}{cc}
\includegraphics[width=0.4\linewidth]{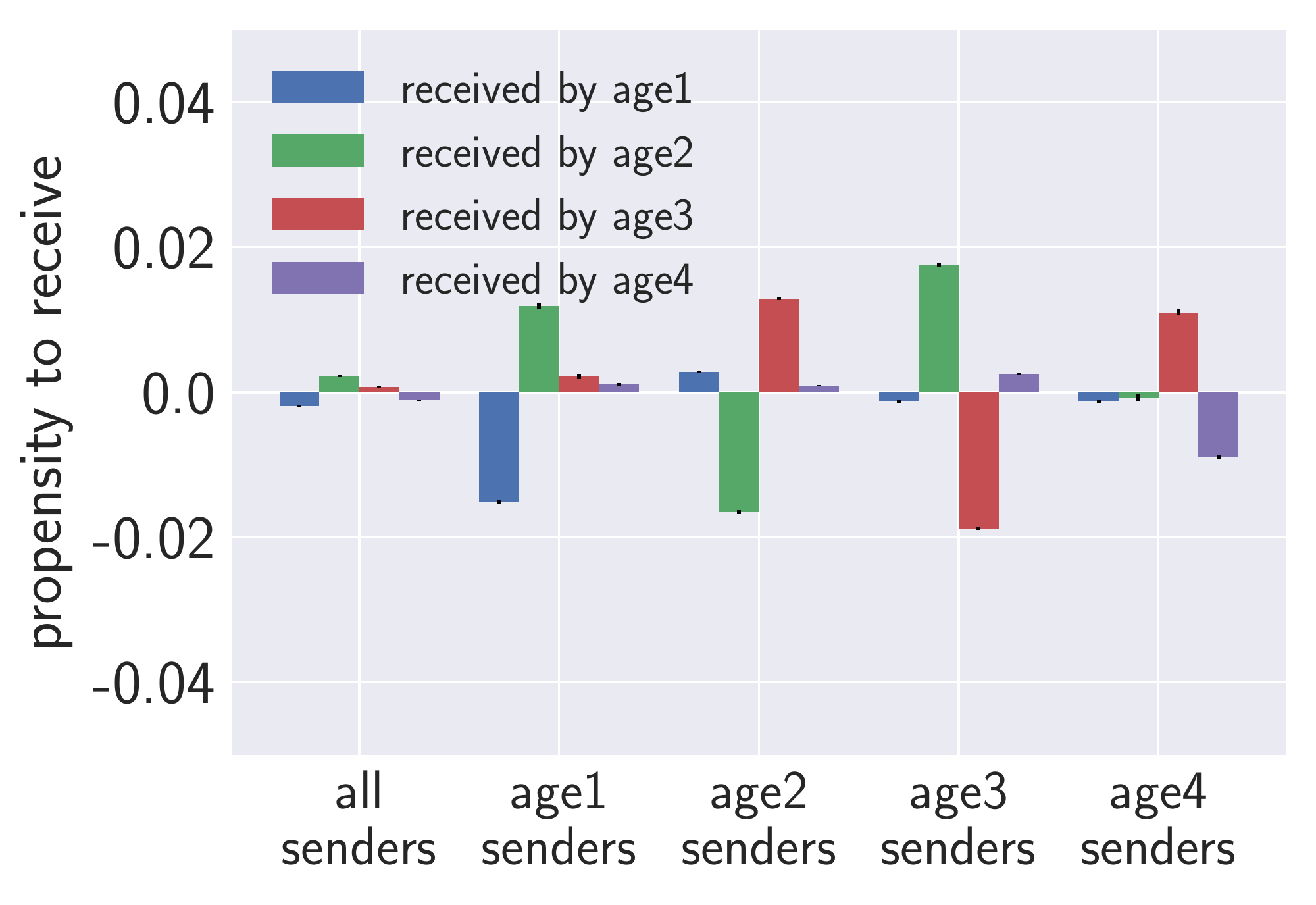} &
\includegraphics[width=0.4\linewidth]{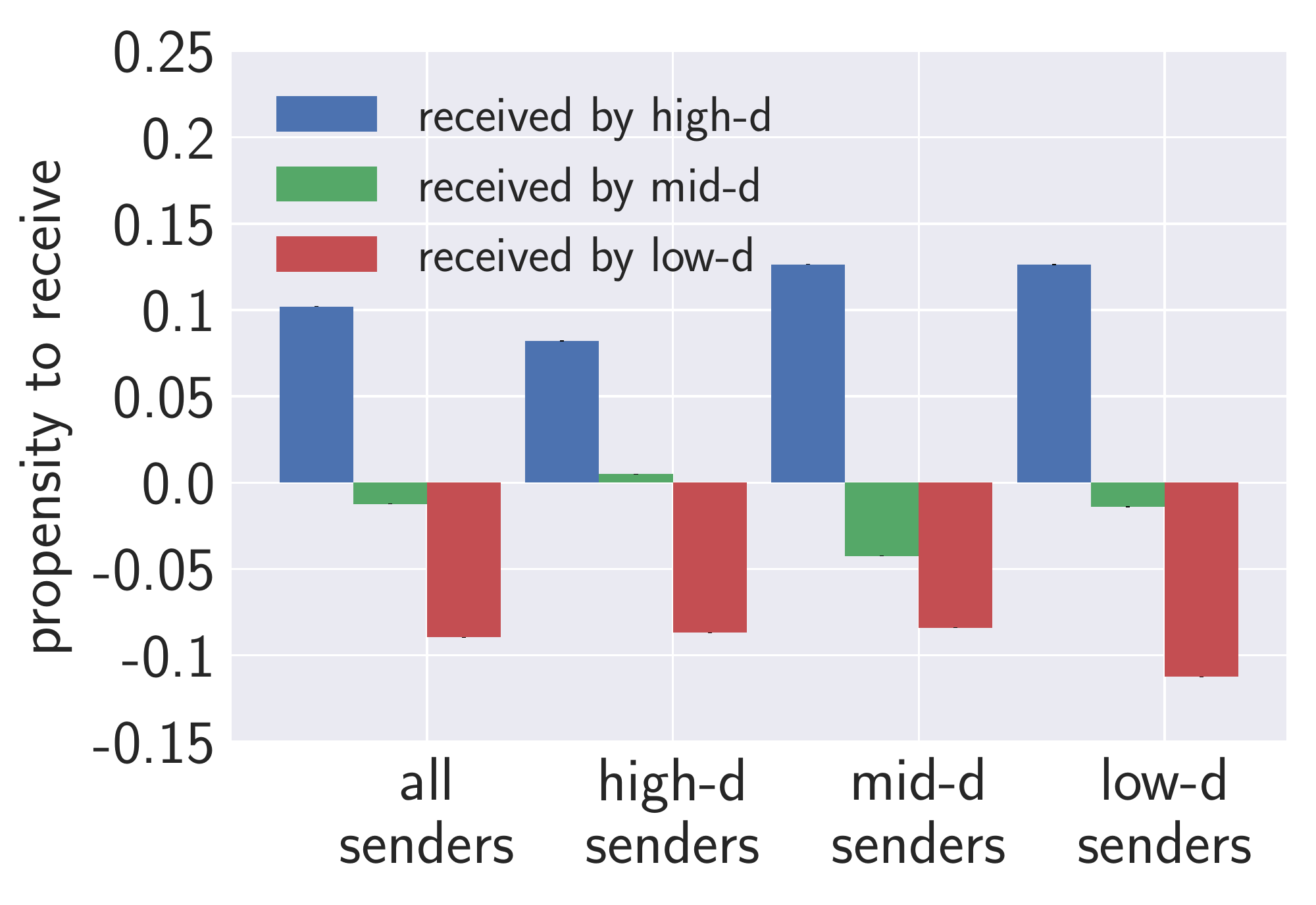} \\
(b) & (c)
\end{tabular}
\caption{Propensity to receive a red packet for given demographic characteristics, represented by the bars. Group-level SEs are shown as blacks ticks on the tops/bottoms of the bars. (a) is gender, (b) age, (c), degree (d), location (e), and strong tie (whether the sender and receiver are friends). We also present the conditions given the sender.}
\label{fig:recv}
\end{figure}

We define the \textit{propensity to receive of characteristic $F$ given condition $C$} and present the propensity conditional on different senders. 
In Figure~\ref{fig:recv}, we depict the propensity to receive of a given characteristic such as gender, age, degree (number of friends), location and whether the sender and receiver are friends.

\begin{itemize}
\item Gender. Females have an approximately $+0.020$ propensity to receive a red packet, while males are less likely to receive red packets. Interestingly, when we consider the genders of red packet senders, we find both genders have a higher propensity to receive a red packet from a person of the opposite gender. Specifically, the propensity of females to receive a red packet from a male is approximately $+0.047$, which is higher than the propensity of males to receive from a female (approximately $+0.01$).  
\item Age. Because we have limited samples of people over age 50, we only consider four groups: \textit{age1}, \textit{age2}, \textit{age3}, \textit{age4}, representing age groups $[10, 20)$, $[20, 30)$, $[30, 40)$, and $[40, 50)$, respectively. On average, the propensity to receive red packets declines with the age when the latter is over 20, indicating that older people tend to be less likely to receive a red packet than a younger person is. 
On average, the propensity to receive red packets declines with the age when the latter is over 20, indicating that older people tend to be less likely to receive a red packet than a younger person is. 
\item Location. As shown in Fig.~\ref{fig:recv}(d), despite the statistical significance, the difference in the propensity to receive across locations is smaller than that for other variables. 
\item In-group degree. In-group degree is the number of friends that a user has in a group. It measures in-group social status. We define three groups: \textit{high-d} (people with in-group degree that is no less than the second tercile), \textit{low-d} (people with in-group degree that is no more than the first tercile) and \textit{mid-d} (the rest). 
We find that a high-degree group member has an approximately $+0.1$ propensity to receive a red packet, regardless of the degree group to which the sender belongs. The propensity decreases as the degree decreases. This is probably because people with high degrees are more avid users of WeChat and are willing to spend more time to wait to receive red packets.
\item Strong ties. Finally, we study whether strong ties differ from weak ties, that is, if one is more likely to receive red packets from friends than from non-friends in a given group. As shown in Figure~\ref{fig:recv}(e), when people are friends, they have an approximately $+0.08$ propensity to receive red packets.
\end{itemize}

\section{PSM details}

Table~\ref{tab:complete} reports the complete regression results.  All covariates are also used as control variables. We report the most significant and interesting controls in the main article.

\begin{table}[!htbp] 
  \caption{Complete Regression Results} 
  \label{tab:complete} 
  \scriptsize \centering 
\begin{tabular}{@{\extracolsep{5pt}}lD{.}{.}{-3} D{.}{.}{-3} D{.}{.}{-3} D{.}{.}{-3} } 
\\[-1.8ex]\hline 
\hline \\[-1.8ex] 
 & \multicolumn{4}{c}{\textit{Dependent variable:}} \\ 
\cline{2-5} 
\\[-1.8ex] & \multicolumn{1}{c}{\#Follower} & \multicolumn{1}{c}{\#FRP} & \multicolumn{1}{c}{Session Money} & \multicolumn{1}{c}{\#Sender’s New Friends} \\ 
\\[-1.8ex] & \multicolumn{1}{c}{(1)} & \multicolumn{1}{c}{(2)} & \multicolumn{1}{c}{(3)} & \multicolumn{1}{c}{(4)}\\ 
\hline \\[-1.8ex] 
 treatment & 0.295^{***} & 0.415^{***} & 52.632^{***} & 0.009^{***} \\ 
  & (0.010) & (0.024) & (1.490) & (0.003) \\ 
  is festival & 0.185^{***} & 0.224^{***} & 6.479^{*} & -0.006 \\ 
  & (0.025) & (0.057) & (3.625) & (0.008) \\ 
  is weekday & -0.094^{***} & -0.078^{***} & -4.862^{***} & -0.002 \\ 
  & (0.011) & (0.025) & (1.602) & (0.003) \\ 
  hour & 0.021^{***} & 0.038^{***} & 1.191^{***} & 0.0001 \\ 
  & (0.001) & (0.002) & (0.118) & (0.0002) \\ 
  sendtime & 0.00000^{***} & 0.00000^{***} & 0.00000^{***} & -0.000 \\ 
  & (0.000) & (0.000) & (0.00000) & (0.000) \\ 
  RPnum & 0.006^{***} & 0.006^{***} & 0.218^{***} & 0.0003^{***} \\ 
  & (0.0004) & (0.001) & (0.055) & (0.0001) \\ 
  latitude & 0.002^{**} & 0.001 & 0.303^{***} & -0.00002 \\ 
  & (0.001) & (0.002) & (0.105) & (0.0002) \\ 
  longitude & -0.001^{*} & -0.0002 & -0.189 & 0.0004 \\ 
  & (0.001) & (0.002) & (0.118) & (0.0002) \\ 
  age & 0.007^{***} & 0.008^{***} & 0.604^{***} & -0.001^{***} \\ 
  & (0.001) & (0.002) & (0.101) & (0.0002) \\ 
  female & 0.089^{***} & 0.119^{***} & 9.683^{***} & 0.002 \\ 
  & (0.013) & (0.029) & (1.815) & (0.004) \\ 
  degree & -0.145^{***} & -0.388^{***} & -12.416^{***} & -0.039^{***} \\ 
  & (0.028) & (0.065) & (4.111) & (0.009) \\ 
  activity\_1 & 0.0002^{***} & 0.0004^{***} & 0.056^{***} & 0.00002 \\ 
  & (0.0001) & (0.0001) & (0.008) & (0.00002) \\ 
  activity\_2 & -0.00003^{**} & -0.0001^{***} & -0.003 & -0.00000 \\ 
  & (0.00001) & (0.00003) & (0.002) & (0.00000) \\ 
  female ratio & -0.130^{***} & -0.218^{***} & -25.545^{***} & 0.001 \\ 
  & (0.026) & (0.059) & (3.699) & (0.008) \\ 
  age entropy & 0.099^{***} & 0.178^{***} & 6.618^{***} & 0.001 \\ 
  & (0.010) & (0.024) & (1.493) & (0.003) \\ 
  province entropy & 0.020^{**} & 0.038^{**} & 3.377^{***} & 0.007^{***} \\ 
  & (0.008) & (0.018) & (1.163) & (0.002) \\ 
  density & -0.267^{***} & -0.301^{***} & 14.506^{***} & -0.013 \\ 
  & (0.030) & (0.068) & (4.307) & (0.009) \\ 
  member count & 0.0004^{***} & 0.002^{***} & 0.015 & 0.001^{***} \\ 
  & (0.0001) & (0.0002) & (0.012) & (0.00002) \\ 
  Constant & -51.669^{***} & -88.131^{***} & -3,267.126^{***} & 0.870 \\ 
  & (2.349) & (5.396) & (340.545) & (0.716) \\ 
 \hline \\[-1.8ex] 
Observations & \multicolumn{1}{c}{108,300} & \multicolumn{1}{c}{108,300} & \multicolumn{1}{c}{108,300} & \multicolumn{1}{c}{108,300} \\ 
R$^{2}$ & \multicolumn{1}{c}{0.036} & \multicolumn{1}{c}{0.019} & \multicolumn{1}{c}{0.016} & \multicolumn{1}{c}{0.051} \\ 
Adjusted R$^{2}$ & \multicolumn{1}{c}{0.036} & \multicolumn{1}{c}{0.019} & \multicolumn{1}{c}{0.016} & \multicolumn{1}{c}{0.051} \\ 
\hline 
\hline \\[-1.8ex] 
\textit{Note:}  & \multicolumn{4}{r}{$^{*}$p$<$0.1; $^{**}$p$<$0.05; $^{***}$p$<$0.01} \\ 
\end{tabular} 
\end{table} 

\end{appendices}

\end{document}